\documentclass[]{article}

\usepackage{arxiv}

\usepackage[utf8]{inputenc} 
\usepackage[T1]{fontenc}    
\usepackage[numbers]{natbib}
\usepackage{hyperref}       
\usepackage{url}            
\usepackage{booktabs}       
\usepackage{amsfonts}       
\usepackage{nicefrac}       
\usepackage{microtype}      
\usepackage{lipsum}
\usepackage{comment}
\usepackage{lmodern}
\usepackage{graphicx}
\usepackage{amsmath}
\usepackage{listings}
\usepackage{subcaption}

\usepackage{xcolor} 
\usepackage{caption}
\usepackage{multicol}
\usepackage{float}
\usepackage{sectsty}

\definecolor{codegreen}{rgb}{0,0.6,0}
\definecolor{codegray}{rgb}{0.5,0.5,0.5}
\definecolor{codepurple}{rgb}{0.58,0,0.82}
\definecolor{backcolour}{rgb}{0.95,0.95,0.92}

\lstdefinestyle{mystyle}{
    backgroundcolor=\color{backcolour},   
    commentstyle=\color{codegreen},
    keywordstyle=\color{magenta},
    numberstyle=\tiny\color{codegray},
    stringstyle=\color{codepurple},
    basicstyle=\ttfamily\footnotesize,
    breakatwhitespace=false,         
    breaklines=true,                 
    captionpos=b,                    
    keepspaces=true,                 
    numbers=left,                    
    numbersep=5pt,                  
    showspaces=false,                
    showstringspaces=false,
    showtabs=false,                  
    tabsize=2
}

\sectionfont{\large}
\usepackage{enumitem}
\setlist{leftmargin=5.5mm}
\usepackage{physics}
\usepackage[toc]{appendix}
\usepackage{titlesec}
\titlespacing*{\section}
{0pt}{20pt}{0pt}
\titlespacing*{\subsection}
{0pt}{20pt}{0pt}
\titlespacing*{\subsubsection}
{0pt}{5pt}{0pt}
\title{Towards graph classification with Gaussian Boson Sampling by embedding graphs on the X8 photonic chip}
\author{
  Edgard Pierre\\
  Prevision.io\\
  Paris, France\\
  \texttt{edgard.pierre@prevision.io}\\
  \And
  Michel Nowak\\
  Prevision.io\\
  Paris, France\\
  \texttt{michel.nowak@prevision.io}\\  
}
\begin{document}
\maketitle
\begin{abstract}
Photonics chips on which one can perform Gaussian Boson Sampling have become accessible on the cloud, in particular the X8 chip of Xanadu. In this technical report, we study its potential use as a first step towards graph classification on quantum devices. In order to achieve this goal, we study the generated samples of the graph embedding method which leads to feature vectors. This is done on a restricted class of unweighted, undirected and loop-free graphs. Hardware constraints are matched to properties of graphs that can be encoded. We report experiments on the X8 chip as well as comparisons to numerical simulations on a classical computer and analytical solutions. We conclude this technical report by trying to take photon loss into account and explain the observed results accordingly.
\end{abstract}

\section{Introduction}
\label{sec:introduction}
Gaussian Boson Sampling (GBS) is currently one of the most promising near term quantum computer. It recently achieved quantum supremacy~\citep{zhong2020quantum}. This year, the company \href{https://www.xanadu.ai/}{Xanadu} released an 8 modes compact GBS photonic chip~\citep{Arrazola2021}. One of their work consisted in studying the possibility of embedding graphs into the chip~\citep{Bromley_2020} in order to perform various operations such as graph isomorphism~\citep{2018arXiv181010644B} and graph similarity~\citep{PhysRevA.101.032314}. At the same time, improvements on graph neural networks~\citep{xu2019powerful} show that graph isomorphism plays a central role in the classification itself. The emergence of their quantum counterparts~\citep{verdon2019quantum} with already proven applications~\citep{Mengoni2021} motivated us to investigate how photonics chips could help in this field. A preliminary part of our study was to determine to which extent graphs can be embedded on a photonic device and how one can derive isomorphism properties from it.

We will thus begin this technical report by describing the hypothesis that we make on the graphs we wish to embed on the chip and match them to the hardware constraints in section~\ref{sec:poss_graphs}. We will then group these possible graphs into categories that are of interest for us, namely, isomorphic groups in section~\ref{sec:graphchar}. At this point, we will be able to run numerical simulations and real calculations on the hardware (section~\ref{sec:GBS}). Finally, we will extract the feature vectors and compare them to expected analytical calculations in section~\ref{sec:fv}.

\begin{figure}[h]
    \centering
    \includegraphics[width=8cm]{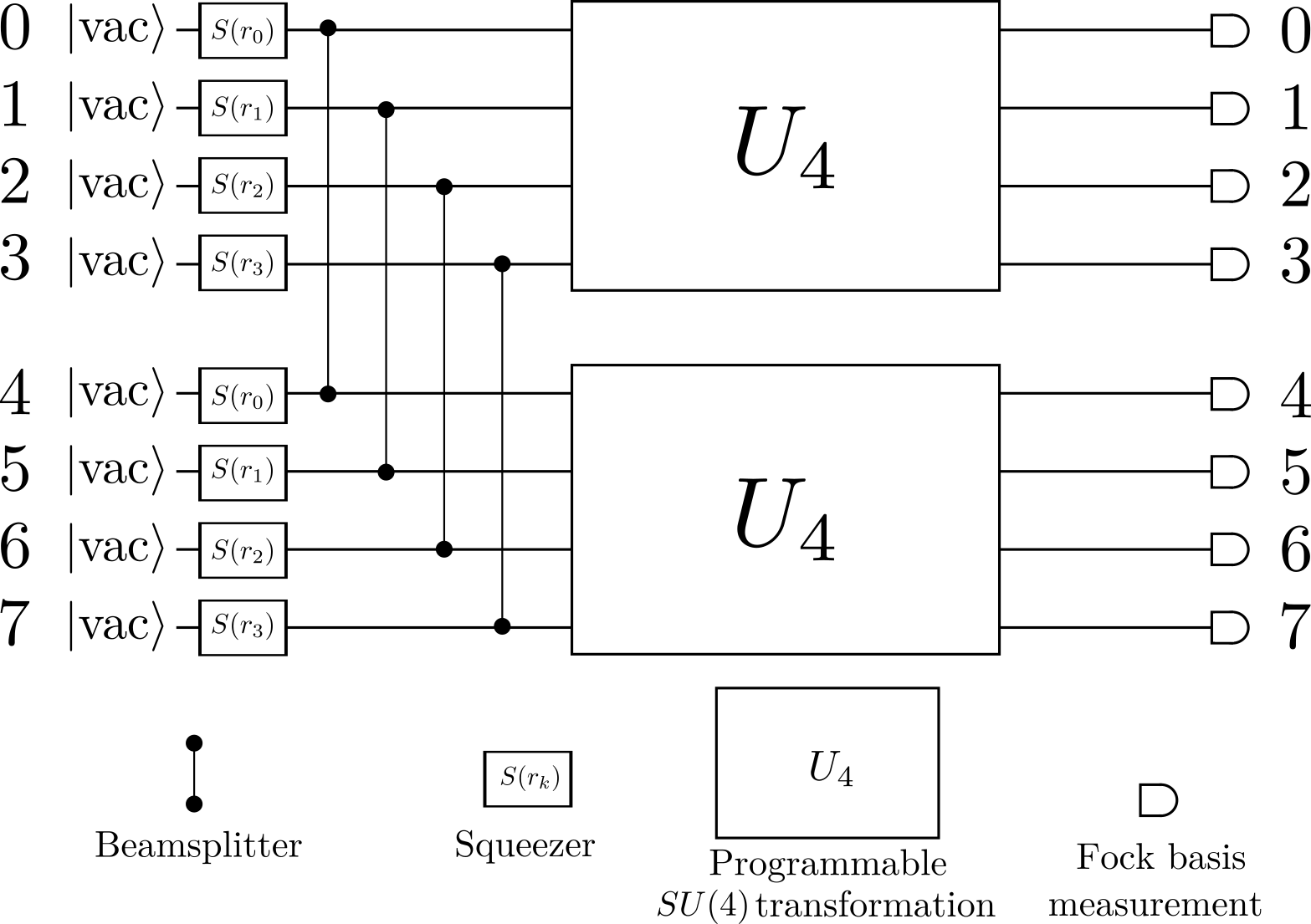}
    \caption{X8 chip circuit (taken from~\citep{Arrazola2021})}
    \label{fig:x8_chip_circuit}
\end{figure}

In order to prepare for the core of the work, we quickly present the X8. Figure~\ref{fig:x8_chip_circuit} shows the circuit that one can implement on this hardware. It is composed of eight modes, each starting in the vacuum state. Then, S2 gates are applied between the first 4 modes (called the signal modes) and the last 4 modes (the idler modes). Subsequently, a unitary transformation is applied on each of the two groups of modes. Finally, the photons are detected on the output channel. The measurement step can be done by counting the number of photons in each mode. We will further refer to this method as the event counting method. Detectors can also be triggered as soon as one photon reaches the output channel. This method will be referred to as the threshold method. 

\section{Possible graph embedding on the X8 chip.}\label{sec:poss_graphs}
We wish to encode graphs on the X8 hardware in order to subsequently derive informative features and solve Machine Learning tasks. The method used for the embedding can only implement graphs with a number of nodes equal to the number of modes of the device and needs to be fed with the associated adjacency matrix $A$. We are thus restricted to working with graphs containing eight nodes. We now present the initial hypothesis that we assume from the graphs we are interested in and map hardware constraints to the possible graphs that can be encoded. We are interested in graphs that:
\begin{itemize}
    \item are unweighted (the matrix elements are 0 or 1)
    \item are undirected (the matrix is symmetric)
    \item do not have self loops (the diagonal of the matrix is 0).
\end{itemize}
As a result one can write the adjacency matrix with 28 independent parameters (which corresponds to $2^{28}\sim 2.68\times 10^{8}$ possible graphs).

\[A =
\begin{pmatrix}
0        & \delta_0    & \delta_1    & \delta_2    & \delta_3    & \delta _4   & \delta_5 & \delta_6      \\
\delta_0 & 0           & \delta_7    & \delta_8    & \delta_9    & \delta_{10} & \delta_{11} & \delta_{12}\\
\delta_1 & \delta_7    & 0           & \delta_{13} & \delta_{14} & \delta_{15} & \delta_{16} & \delta_{17}\\
\delta_2 & \delta_8    & \delta_{13} & 0           & \delta_{18} & \delta_{19} & \delta_{20} & \delta_{21}\\
\delta_3 & \delta_9    & \delta_{14} & \delta_{18} & 0           & \delta_{22} & \delta_{23} & \delta_{24}\\
\delta_4 & \delta_{10} & \delta_{15} & \delta_{19} & \delta_{22} & 0           & \delta_{25} & \delta_{26}\\
\delta_5 & \delta_{11} & \delta_{16} & \delta_{20} & \delta_{23} & \delta_{25} & 0           & \delta_{27}\\
\delta_6 & \delta_{12} & \delta_{17} & \delta_{21} & \delta_{24} & \delta_{26} & \delta_{27} & 0          \\
\end{pmatrix}
\]

with $\delta_{i} = 0$ or $1$.

\subsection*{Bipartite graphs}
The X8 chip can only embed bipartite graphs because of its structure. These graphs' adjacency matrices are defined as the following:

\[A =
\begin{pmatrix}
0 & M\\
M^{T} & 0
\end{pmatrix}
\]

with $M$ a given sub-matrix and $M^{T}$ its transpose.
With this configuration, one can see the number of independent parameters is reduced to 16 ($65536$ possible graphs).

\[A =
\begin{pmatrix}
0        & 0        & 0           & 0           & \delta_0    & \delta _1   & \delta_2    & \delta_3   \\
0        & 0        & 0           & 0           & \delta_4    & \delta_5    & \delta_6    & \delta_7   \\
0        & 0        & 0           & 0           & \delta_8    & \delta_9    & \delta_{10} & \delta_{11}\\
0        & 0        & 0           & 0           & \delta_{12} & \delta_{13} & \delta_{14} & \delta_{15}\\
\delta_0 & \delta_4 & \delta_8    & \delta_{12} & 0           & 0           & 0           & 0          \\
\delta_1 & \delta_5 & \delta_9    & \delta_{13} & 0           & 0           & 0           & 0          \\
\delta_2 & \delta_6 & \delta_{10} & \delta_{14} & 0           & 0           & 0           & 0          \\
\delta_3 & \delta_7 & \delta_{11} & \delta_{15} & 0           & 0           & 0           & 0          \\
\end{pmatrix}
\]

\subsection*{Signal and idler mode constraint}

The hardware constraints regarding the connections between the first four modes (the signal modes) of the X8 chip and the four last (the idler modes) also apply a new condition on the sub-matrix M: it must be symmetric. One can then reduce the number of independent parameters to 10 ($1024$ possible graphs).

\[M =
\begin{pmatrix}
\delta_0 & \delta_1 & \delta_2 & \delta_3\\
\delta_1 & \delta_4 & \delta_5 & \delta_6\\
\delta_2 & \delta_5 & \delta_7 & \delta_8\\
\delta_3 & \delta_6 & \delta_8 & \delta_9\\
\end{pmatrix}
\]

Until the end of this document, we will name the graphs as "$\delta_0\delta_1\delta_2\delta_3\delta_4\delta_5\delta_6\delta_7\delta_8\delta_9$" (see Tab.\ref{table:graph_list}).

\subsection*{Normalization constraint}

A last constraint on the adjacency matrix elements comes from the embedding method. Indeed, we use the \href{https://strawberryfields.ai}{Strawberry Fields} (sf) library in order communicate with the X8 chip and embed the graphs. The \verb|BipartiteGraphEmbed| class from the \verb|sf.ops| module requires the use of a \verb|m| mean photon per mode normalization parameter in order to be sure that the singular values are scaled such that all Sgates have squeezing parameter \verb|r=1|. Whatever the \verb|m| parameter is, the only configurations embeddable are the ones where the \verb|r| parameters for all the four signal modes are identical and or equals to zero. 
This gives the following constraint on $M$:

\[M=U \mathrm{diag}(\tanh r_i )U^T\]

where $r_i$ is the squeezing parameter on the $i$-th pair of modes and $U$ the interferometer used to prepare the state.

This leaves 75 graphs over the 1024 theoretically embeddable graphs fulfilling these conditions. We noticed that the value of \verb|m| depends on the number of substructures of the graph (called "subgraph" in the next section). For one substructure, we have $m=m_{0}=0.345274461385554870545$. For \verb|n| substructures (up to \verb|n=4|), we have $m=n\times m_{0}$.

\section{Graphs characterization and isomorphism.}\label{sec:graphchar}
Now that we have put constraints on the possible graphs that are of interest to us, we characterize them into groups. The terminology used in the coming listing is used assuming the reader has some basics in the graph theory.

Despite being able to embed 75 different graphs into the X8 chip, we noticed most of them were isomorphic. It is possible to group them into 10 categories:
\begin{itemize}
    \item the graph contains a $K_2$ subgraph and 6 free vertices. Called "1K2". $m=m_0$
    \item the graph contains 2 $K_2$ subgraphs and 4 free vertices. Called "2K2". $m=2\times m_0$
    \item the graph contains a $C_4$ subgraph and 4 free vertices. Called "1C4". $m=m_0$
    \item the graph contains 2 $P_3$ subgraphs and 2 free vertices. Called "2P3". $m=2\times m_0$
    \item the graph contains 3 $K_2$ subgraphs and 2 free vertices. Called "3K2". $m=3\times m_0$
    \item the graph contains a Thomsen $K_{3,3}$ subgraph and 2 free vertices. Called "1K33". $m=m_0$
    \item the graph contains 2 $S_3$ subgraphs and no free vertex. Called "2S3". $m=2\times m_0$
    \item the graph contains 4 $K_2$ subgraphs and no free vertex. Called "4K2". $m=4\times m_0$
    \item the graph contains 2 $C_4$ subgraphs and no free vertex. Called "2C4". $m=2\times m_0$
    \item the graph is a $K_{4,4}$ graph. Called "1K44". $m=m_0$
\end{itemize}
One can note that the only embeddable connected graph is \verb|1111111111|. We will use this graph as an example for the last part of our study (related to errors, see section \ref{sec:fv}).

\begin{figure}[h!]
    \centering
    \begin{subfigure}{0.45\textwidth}
     \centering
     \includegraphics[width=\textwidth]{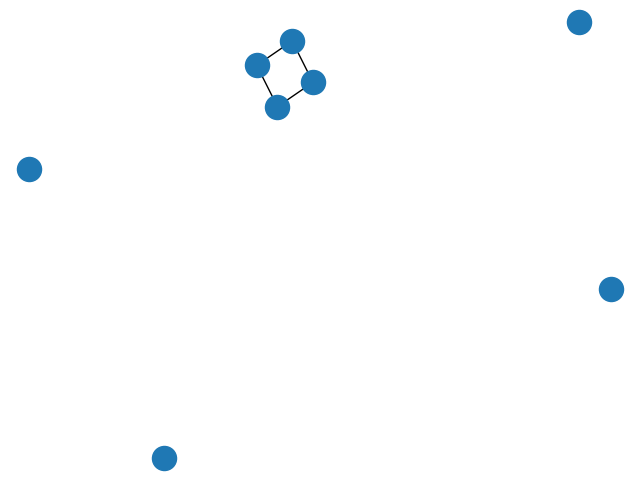}
     \caption{planar view}
     \label{fig:1C4_planar_view}
    \end{subfigure}
    \hfill
    \begin{subfigure}{0.45\textwidth}
     \centering
     \includegraphics[width=\textwidth]{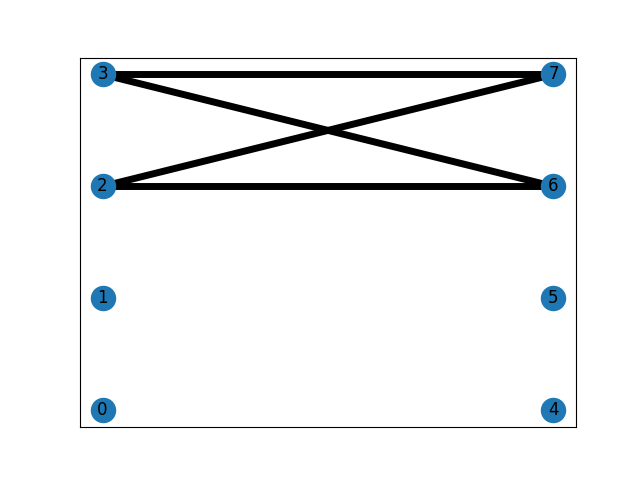}
      \caption{bipartite view}
      \label{fig:1C4_bipartite_view}
    \end{subfigure}
    \caption{1C4 illustrations}
    \label{fig:1C4_illustrationg}
\end{figure}

We illustrate one simple and one complex graph for the reader. The first simple graph is an 8 node graph with only 4 nodes connected to each other in a square fashion as can be viewed in its planar representation in Figure~\ref{fig:1C4_planar_view} or in its bipartite representation in Figure~\ref{fig:1C4_bipartite_view}.

\begin{figure}[h!]
    \begin{subfigure}{0.45\textwidth}
     \centering
     \includegraphics[width=\textwidth]{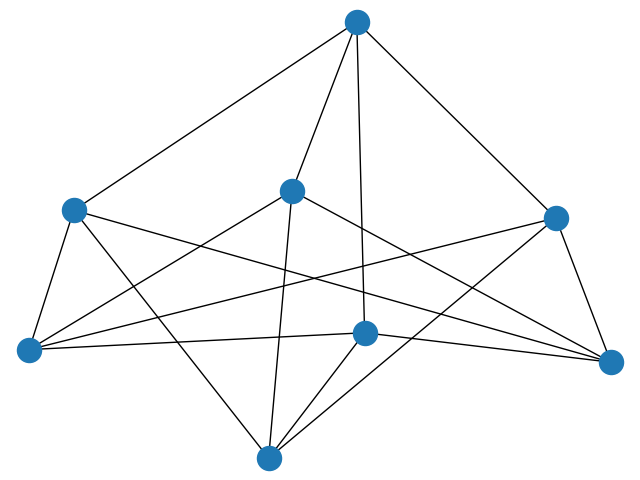}
      \caption{planar view}
      \label{fig:1K44_planar_view}
    \end{subfigure}
    \hfill
    \begin{subfigure}{0.45\textwidth}
     \centering
     \includegraphics[width=\textwidth]{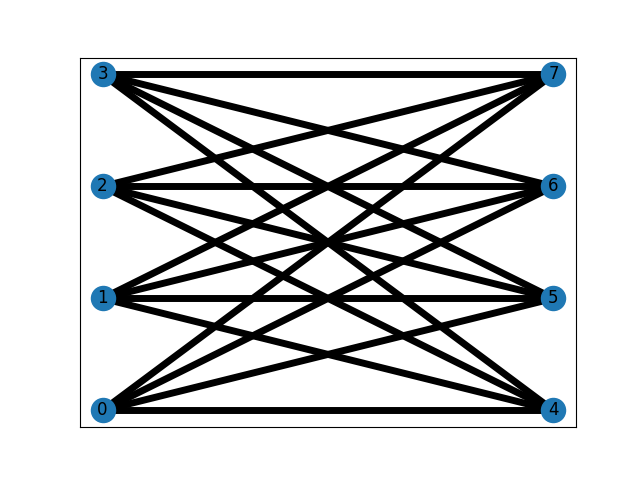}
      \caption{bipartite view}
      \label{fig:1K44_bipartite_view}
    \end{subfigure}
    \label{fig:1K44_illustrations}
    \caption{1K44 illustrations}
\end{figure}

We also illustrate a more complex graph that will be used in later analysis: the 1K44 graph in its planar representation in Figure~\ref{fig:1K44_planar_view} and in its bipartite representation Figure~\ref{fig:1K44_bipartite_view}.

Other graphs representations can be found in Appendix~\ref{sec:appendix}.

Table \ref{table:graph_list} summarizes the different possible embeddable unweighted graphs on the X8 chip. 
\begin{table}[h!]
\centering
 \begin{tabular}{||c c c||} 
 \hline
 Group & Number of graphs & Names \\ [0.5ex] 
 \hline\hline
 1K2 & 4  & 0000000100, 1000000000, 0000100000, 0000000001 \\ 
 \hline
 2K2 & 12 & 1000100000, 1000000100, 1000000001, 0010000000, 0000000010, 0000100100,\\ 
     &    & 0000100001, 0100000000, 0000010000, 0001000000, 0000000101, 0000001000 \\
 \hline
 1C4 & 6  & 1100100000, 0000000111, 1010000100, 1001000001, 0000101001, 0000110100 \\
 \hline
 2P3 & 12 & 0010010000, 0100001000, 0101000000, 0000001010, 0010000010, 0001001000,\\
     &    & 0001000010, 0110000000, 0100010000, 0011000000, 0000010010, 0000011000\\
 \hline
 3K2 & 16 & 0001000100, 1000000010, 1000100100, 1000001000, 0010000001, 0000010001,\\
     &    & 0100000100, 1000010000, 0000100010, 1000100001, 0010100000, 0000100101,\\
     &    & 0000001100, 0100000001, 1000000101, 0001100000 \\ [1ex] 
  \hline
 1K33 & 4 & 1011000111, 1101101001, 1110110100, 0000111111 \\ [1ex] 
  \hline
 2S3 & 4  & 0100011000, 0001001010, 0010010010, 0111000000 \\ [1ex] 
  \hline
 4K2 & 10 & 1000100010, 0100000010, 0010001000, 1000010001, 0001010000, 0010100001,\\
     &    & 1000100101, 0001100100, 1000001100, 0100000101 \\ [1ex] 
  \hline
 2C4 & 6  & 1100100111, 0011011000, 0101010010, 0110001010, 1010101101, 1001110101 \\ [1ex] 
  \hline
 1K44 & 1 & 1111111111 \\ [1ex] 
 \hline
\end{tabular}
\caption{List of all the embeddable unweighted graphs.}
\label{table:graph_list}
\end{table}

\section{Generation of GBS samples and simulated samples}\label{sec:GBS}

We shot 100,000 GBS samples for each of the 75 embeddable configurations. This was done on the same physical device (named \verb|X8_01|). Although this statistics may be insufficient - as in the graph similarity study of~\citep{Arrazola2021}, each configuration used 20 million samples - it was not possible to get such high precision statistics by the time of the study.

We used the following code to generate the GBS outputs. This code was adapted from the Strawberryfields "X8 tutorial" demo available here: \url{https://strawberryfields.ai/photonics/demos/tutorial_X8.html}

\begin{lstlisting}[language=Python]
import strawberryfields as sf
nb_shot = 100000
for A, m in graph_parameters:
    eng = RemoteEngine("X8")
    prog = sf.Program(8)
    with prog.context as q:
        sf.ops.BipartiteGraphEmbed(A, mean_photon_per_mode=m) | q
        sf.ops.MeasureFock() | q
    spec = eng.device_spec
    prog.compile(device=spec)
    run_prog = eng.run(prog, shots=nb_shot)
    GBS_samples = run_prog.samples
\end{lstlisting}
with \verb|A| the adjacency matrix of the embedded graph and \verb|m| the mean photon per mode parameter described in section~\ref{sec:poss_graphs}.

In addition, we also simulated for each configuration 1,000 GBS samples (and not 100,000 because of time constraints since each configuration was about 90 minutes computation time with the \verb|threshold| option off on a Dell XPS15 laptop with a i7-9750H processor and 16 GB RAM) using the same \verb|A| matrix and the same \verb|m| parameter with the \verb|sf.sample.sample| function. The minimalist used code is the following:
\begin{lstlisting}[language=Python]
nb_shot = 1000
for A, m in graph_parameters:
    n_mean = 8*m #m was "per mode", n_mean is for the whole device
    simulated_samples = sf.sample.sample(A, n_mean, nb_shot, threshold=False)
\end{lstlisting}

These sets of GBS samples were saved and used as input for the calculation of feature vectors.

\section{Feature vectors determination: comparison between experimental sampling, simulated sampling and analytical calculation}\label{sec:fv}
We are now able to convert the hardware and simulated samples into "feature vectors" (fv) as described in~\citep{PhysRevA.101.032314} and in the sf tutorial as well, available at this url: \url{https://strawberryfields.ai/photonics/apps/run_tutorial_similarity.html}.
Feature vectors can "be viewed as a mapping to a high-dimensional feature space".

They can be built considering "events" or "orbits". A n-photons event corresponds to n photons detected at the end of the GBS device. Orbits are sub parts of events and are the set of all GBS samples that are equivalent under permutation of the modes. We calculated feature vectors for all the possible events and orbits up to 8 photons. Since most feature vectors components were almost zero for lots of orbits, we first focused on feature vectors from events. Also, since the photons are emitted by pairs, all the odd number of photon events are null under the assumption of a lossless device.

The following minimalist code shows how the feature vectors were calculated on the hardware.

\begin{lstlisting}[language=Python]
event = [2, 4, 6, 8]
mcpm = 8 #max count per mode
for A, m in graph_parameters:
    n_mean = 8*m #m was "per mode", n_mean is for the whole device
    fv_GBS = sf.similarity.feature_vector_events_sampling(GBS_samples, 
                                                          event, mcpm)
    fv_simu = sf.similarity.feature_vector_events_sampling(simulated_samples,
                                                           event, 
                                                           mcpm)
    fv_adj = sf.similarity.feature_vector_events(nx.Graph(A),
                                                 event, 
                                                 mcpm, 
                                                 n_mean)
\end{lstlisting}
\subsection*{Feature vectors comparison}

\begin{figure}[htp]
    \centering
    \includegraphics[width=16cm]{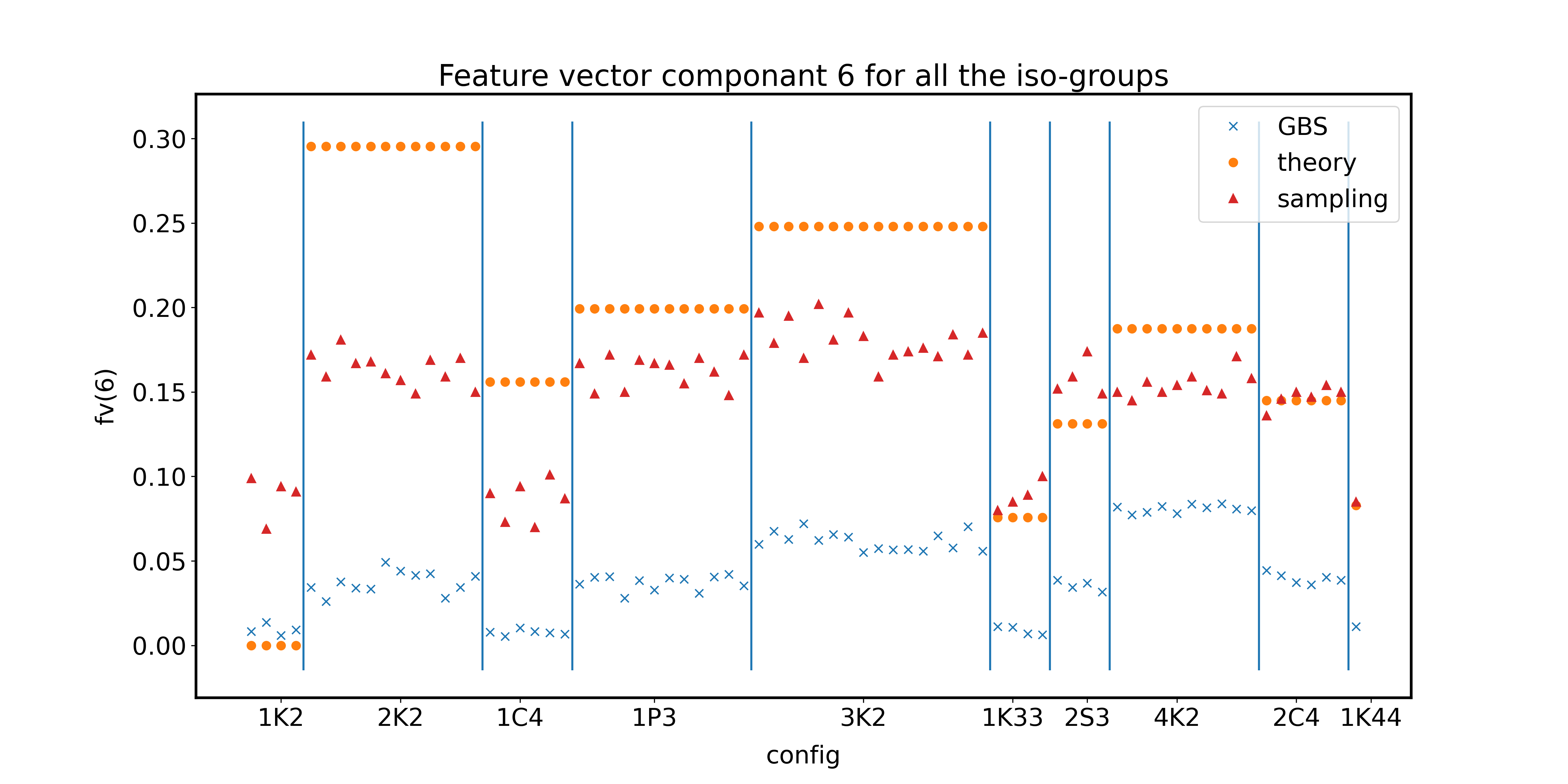}
    \caption{Feature vector for event 6}
    \label{fig:fv6}
\end{figure}

Figure \ref{fig:fv6} shows the third component of the feature vector $f_{k=(2,4,6,8),n_{max=8}}$ for all the 75 embeddable graphs, calculated by three methods:
\begin{itemize}
    \item GBS: using the experimental samples
    \item sampling: using the simulated samples
    \item theory: using the adjacency matrix (analytical calculation).
\end{itemize}

Configurations are sorted on the $x$ axis by isomorphic category.
We notice that each graph of a given isomorphic category has the same theoretical feature vector, in agreement with theory. Furthermore, each graph of a given isomorphic category has a value around an average with a given dispersion (up to 15\% for the GBS hardware sampled fv components and 25\% for the classically simulated sampling fv components). Dispersion is bigger for the simulated ones since the number of simulated shots was very small (1,000 per graph). This low statistics on the simulated sampling prevents us to get relevant conclusions on this comparison.

Nevertheless, we observe the GBS components always have smaller values for events of 6 photons. We expect this to be due to photon loss, and try to verify this in the following section.

\subsection*{Taking photon loss into account}
These comparisons were made under the assumption the systems were lossless. This is not the case for the real GBS device. In order to check this aspect, we considered a configuration (graph \verb|1111111111|) where theory and simulated sampling were in agreement but not the samples from the device and added a loss parameter for the theory. Indeed, the \verb|sf.similarity.feature_vector_events| function has a "loss" argument (set to zero as default) simulating photon losses.

\begin{figure}[htp]
    \centering
    \includegraphics[width=16cm]{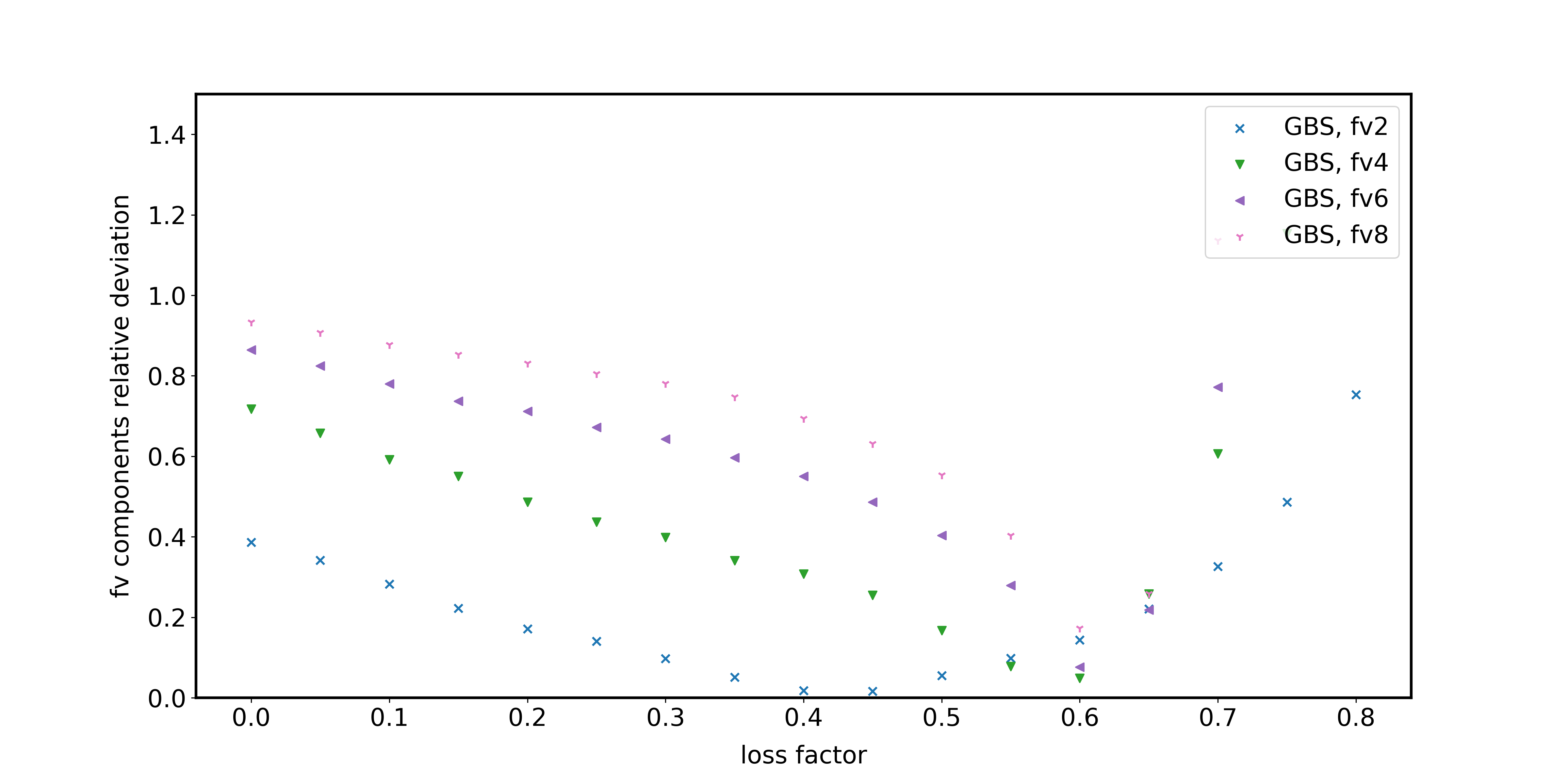}
    \caption{Feature vectors components relative deviation as a function of the loss factor for the 1111111111 graph}
    \label{fig:loss}
\end{figure}

Figure~\ref{fig:loss} shows the relative deviation from theory of the components of the feature vector with respect to different loss factors. One can see that this loss factor must be set to about 45\% in order to match the first component and about 60\% for the three others components. A constant loss factor cannot make an agreement between the theoretical feature vector and the one calculated from the GBS samples. Nevertheless, the loss factor must be set to a pretty large value (close to 60\% photon loss) in order to get one component matching with the theory, which is in agreement with the loss factor estimated in~\citep{Arrazola2021} (about 8 dB integrated loss).

\subsection*{Working with orbits instead of events}
Features vectors made from events do not seem to be the most efficient way to estimate if the GBS device can separate isomorphic groups. Since a proof of graph similarity has already been made using the X8 chip~\citep{Arrazola2021}, we decided to use the same strategy: considering three different orbits, including odd ones (which can happen since the device has loss. We thus use the loss as a feature!) and representing graphs in this 3D space. The chosen orbits were the same as the original paper: [1, 1, 1], [1, 1, 1, 1], and [2, 1, 1].

\begin{figure}[htp]
    \centering
    \includegraphics[width=16cm]{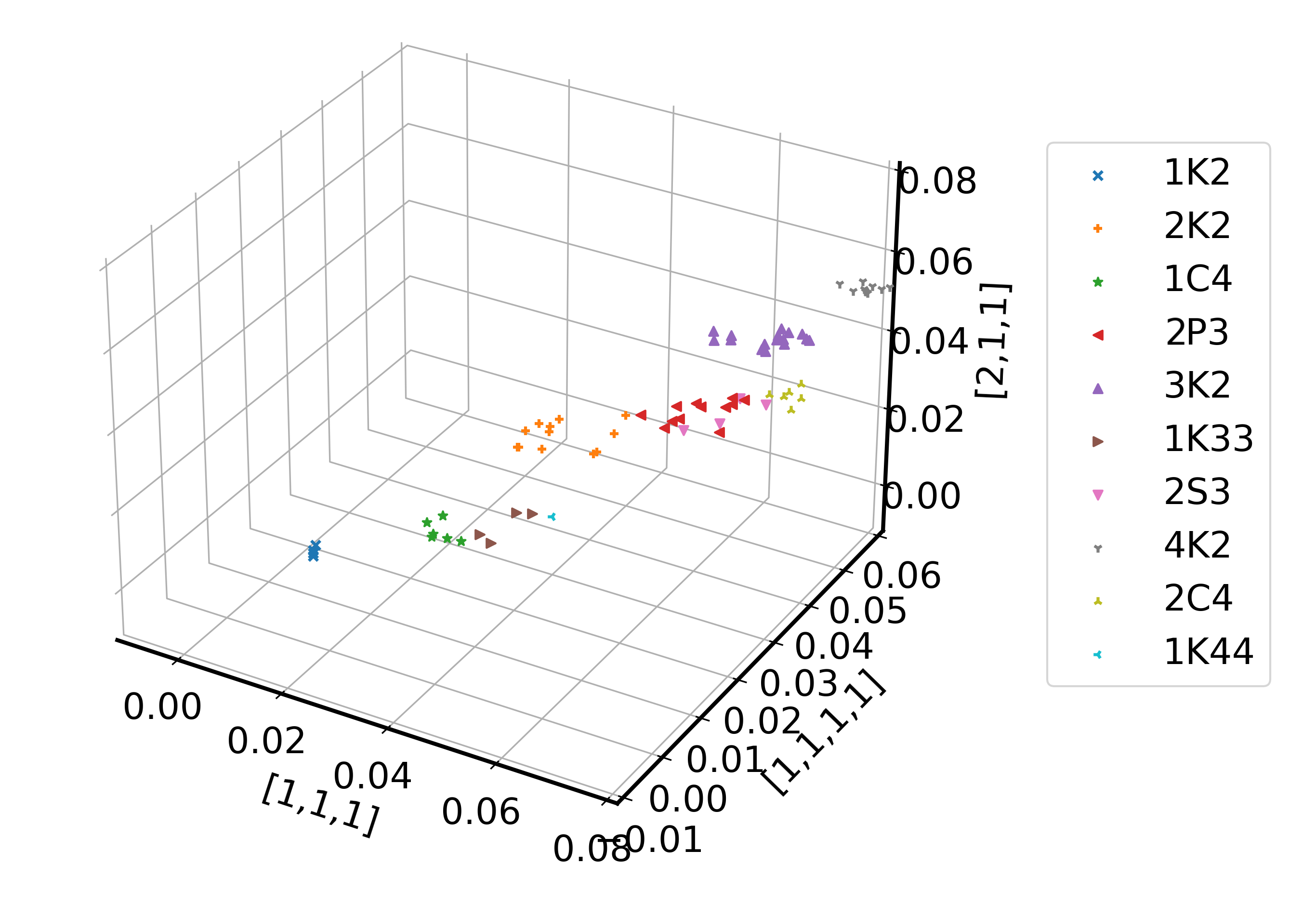}
    \caption{Feature vectors made from orbits. The components of the vectors are probabilities for the orbits [1, 1, 1], [1, 1, 1, 1], and [2, 1, 1].}
    \label{fig:fv_exp}
\end{figure}

Results are shown on Fig.\ref{fig:fv_exp}. Each isomorphic category is clustered and only two (2P3 and 2S3) show overlaps. This is a promising result regarding the loss of the device. The result could certainly be improved with a better statistics.

\section{Conclusion}\label{sec:conclusion}

This technical note summarises a study made about unweighted graphs. We first characterized which unweighted graphs can be embedded into the X8 chip. We noticed 75 unweighted bipartite graphs can be embedded, distributed into 10 different isomorphic categories. Each graph was embedded into the X8 chip and GBS samples were collected. Similar samples were also simulated with a lower statistics. For all of these configurations, feature vectors from events and orbits were calculated. Selecting relevant orbits (including odd ones) allows the classification of the isomorphic categories to a certain extent despite the 8 dB integrated loss of the device.

\section*{Acknowledgements}\label{sec:aknowledgements}

The authors want to thanks Xanadu for letting them use their device. A special attention for Thomas R. Bromley and Juan Miguel Arrazola for their relevant comments.

\bibliographystyle{unsrt}
\bibliography{}
\newpage
\appendix
\section{Appendix: the 10 different graphs}\label{sec:appendix}

The coming plots show the ten different graphs, depicted in two ways: left is by avoiding edge crossing and right is in a more usual bipartite way.

\begin{figure}[htp]
     \centering
     \begin{subfigure}[b]{0.45\textwidth}
         \centering
         \includegraphics[width=\textwidth]{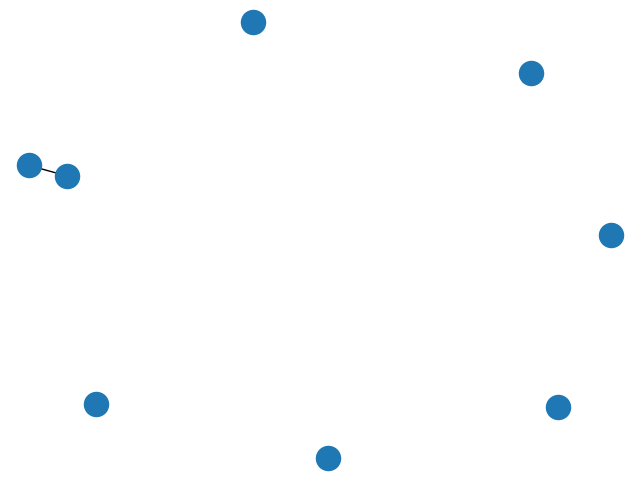}
     \end{subfigure}
     \hfill
     \begin{subfigure}[b]{0.45\textwidth}
         \centering
         \includegraphics[width=\textwidth]{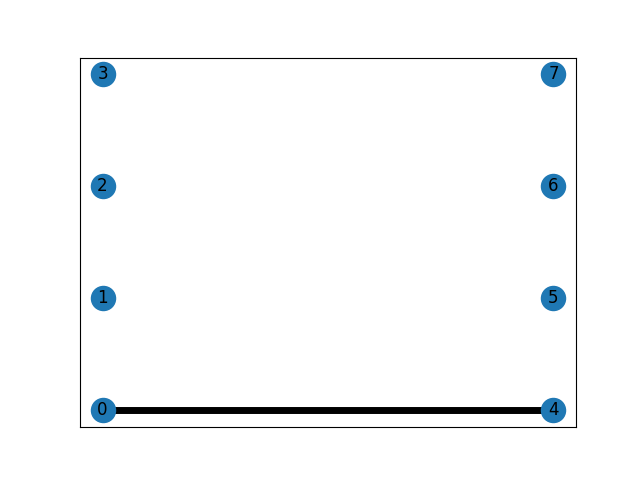}
     \end{subfigure}
        \caption{The "1K2" graph}
        \label{fig:1K2}
\end{figure}

\begin{figure}[htp]
     \centering
     \begin{subfigure}[b]{0.45\textwidth}
         \centering
         \includegraphics[width=\textwidth]{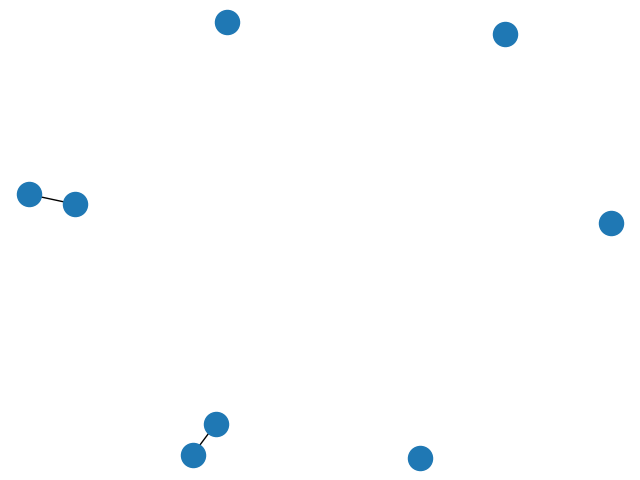}
     \end{subfigure}
     \hfill
     \begin{subfigure}[b]{0.45\textwidth}
         \centering
         \includegraphics[width=\textwidth]{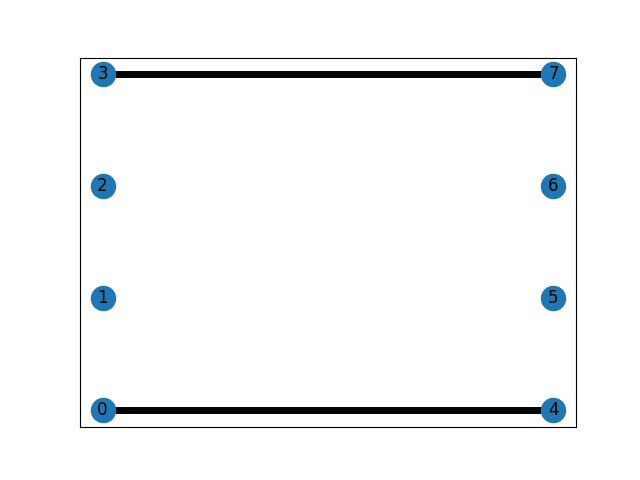}
     \end{subfigure}
        \caption{The "2K2" graph}
        \label{fig:2K2}
\end{figure}

\begin{figure}[htp]
     \centering
     \begin{subfigure}[b]{0.45\textwidth}
         \centering
         \includegraphics[width=\textwidth]{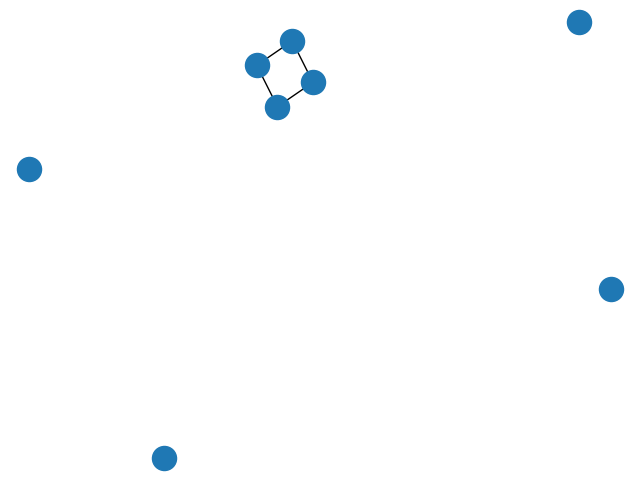}
     \end{subfigure}
     \hfill
     \begin{subfigure}[b]{0.45\textwidth}
         \centering
         \includegraphics[width=\textwidth]{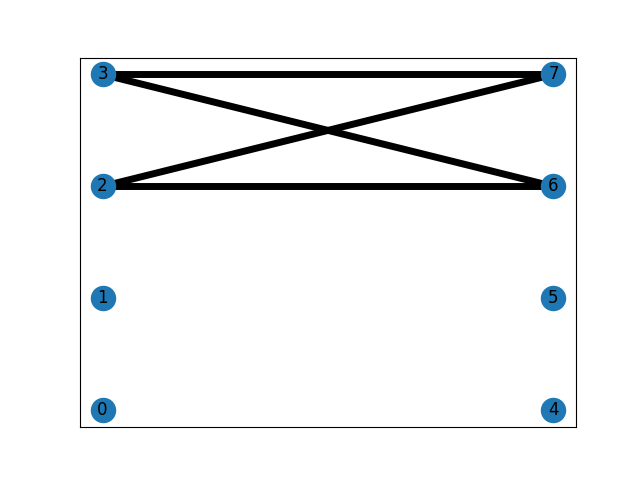}
     \end{subfigure}
        \caption{The "1C4" graph}
        \label{fig:1C4}
\end{figure}

\begin{figure}[htp]
     \centering
     \begin{subfigure}[b]{0.45\textwidth}
         \centering
         \includegraphics[width=\textwidth]{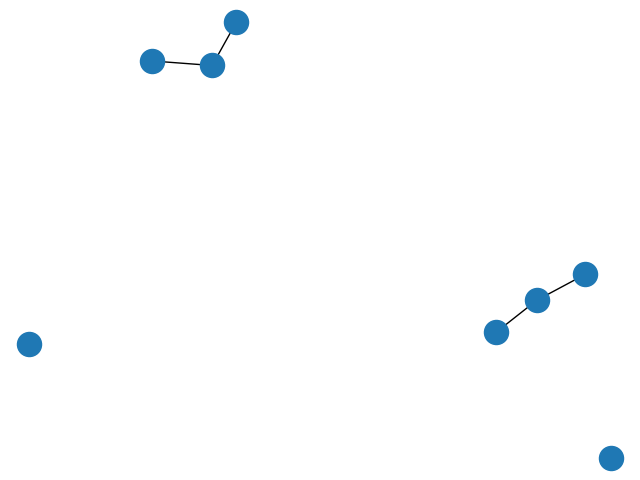}
     \end{subfigure}
     \hfill
     \begin{subfigure}[b]{0.45\textwidth}
         \centering
         \includegraphics[width=\textwidth]{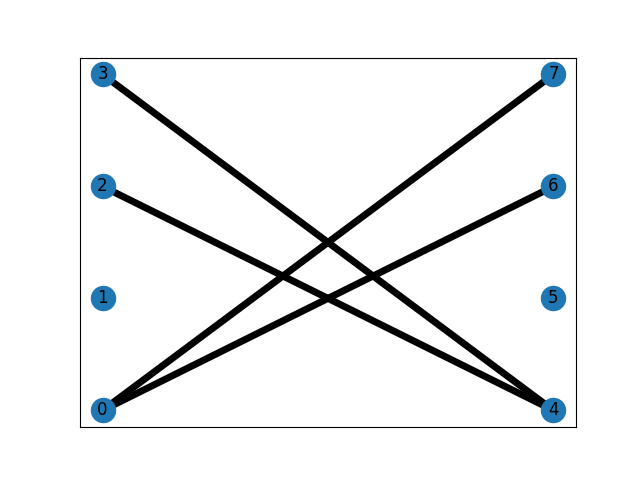}
     \end{subfigure}
        \caption{The "2P3" graph}
        \label{fig:2P3}
\end{figure}

\begin{figure}[htp]
     \centering
     \begin{subfigure}[b]{0.45\textwidth}
         \centering
         \includegraphics[width=\textwidth]{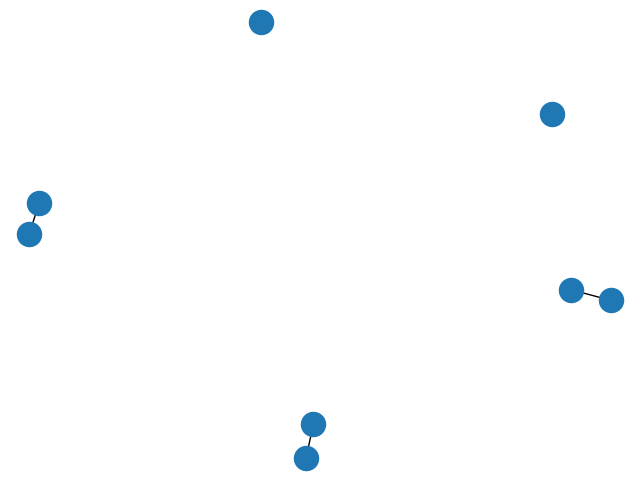}
     \end{subfigure}
     \hfill
     \begin{subfigure}[b]{0.45\textwidth}
         \centering
         \includegraphics[width=\textwidth]{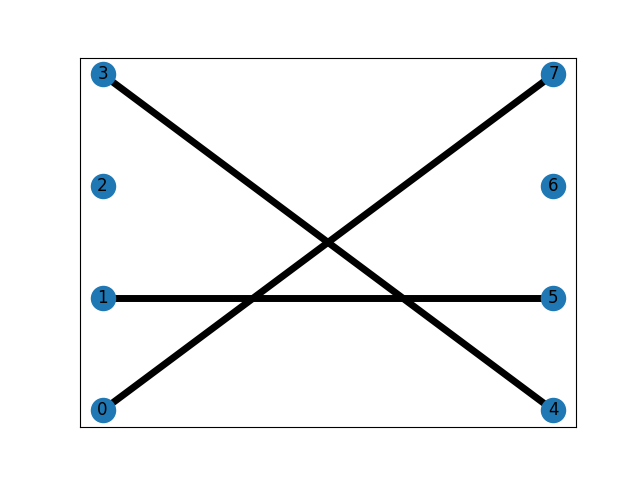}
     \end{subfigure}
        \caption{The "3K2" graph}
        \label{fig:3K2}
\end{figure}

\begin{figure}[htp]
     \centering
     \begin{subfigure}[b]{0.45\textwidth}
         \centering
         \includegraphics[width=\textwidth]{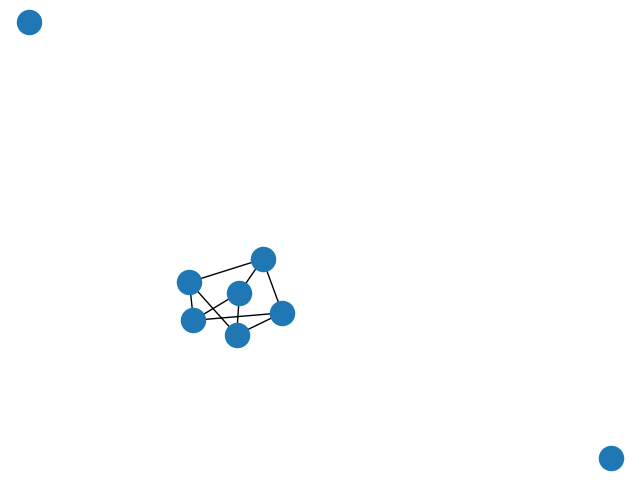}
     \end{subfigure}
     \hfill
     \begin{subfigure}[b]{0.45\textwidth}
         \centering
         \includegraphics[width=\textwidth]{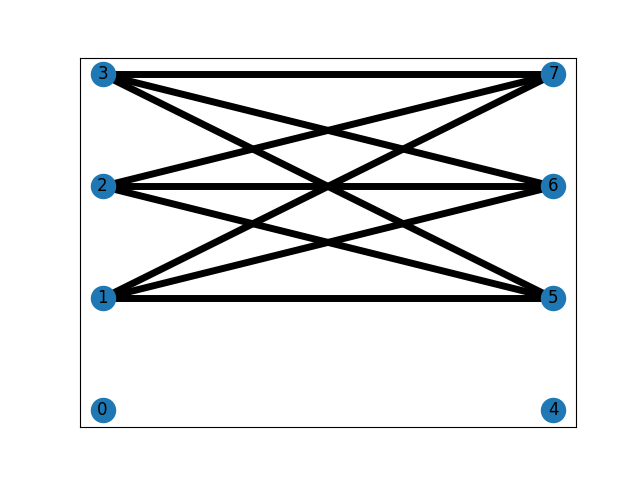}
     \end{subfigure}
        \caption{The "1K33" graph}
        \label{fig:1K33}
\end{figure}

\begin{figure}[htp]
     \centering
     \begin{subfigure}[b]{0.45\textwidth}
         \centering
         \includegraphics[width=\textwidth]{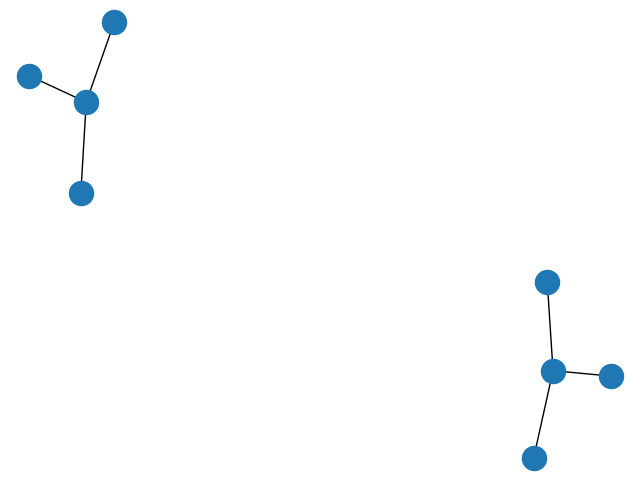}
     \end{subfigure}
     \hfill
     \begin{subfigure}[b]{0.45\textwidth}
         \centering
         \includegraphics[width=\textwidth]{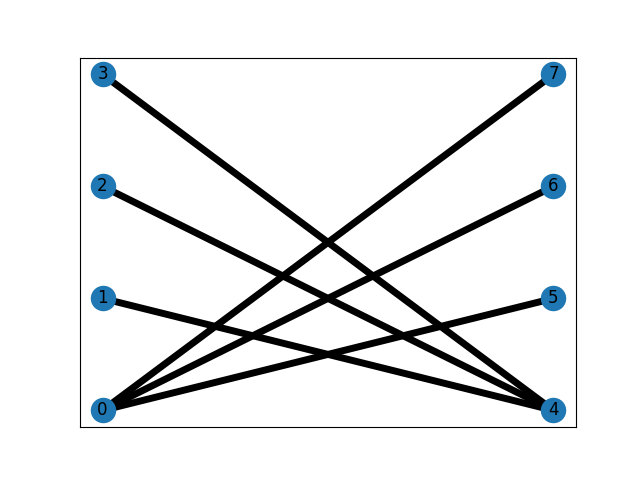}
     \end{subfigure}
        \caption{The "2S3" graph}
        \label{fig:2S3}
\end{figure}

\begin{figure}[htp]
     \centering
     \begin{subfigure}[b]{0.45\textwidth}
         \centering
         \includegraphics[width=\textwidth]{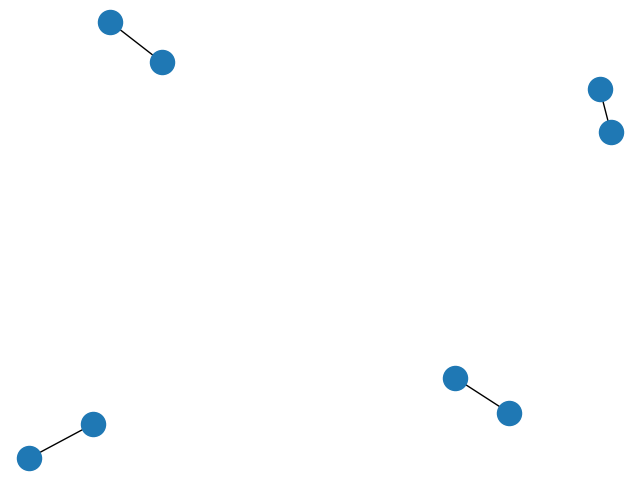}
     \end{subfigure}
     \hfill
     \begin{subfigure}[b]{0.45\textwidth}
         \centering
         \includegraphics[width=\textwidth]{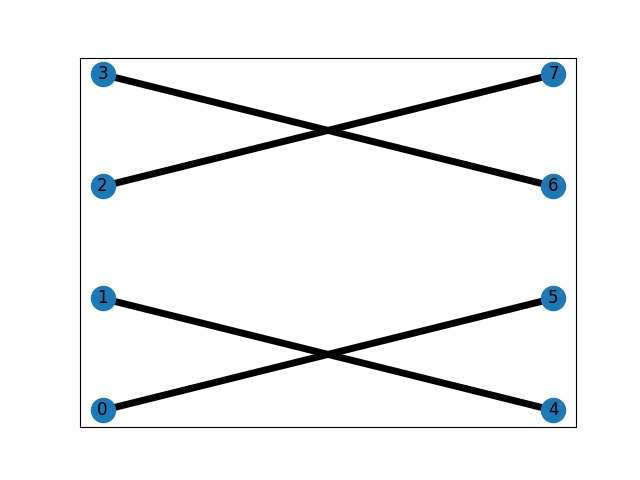}
     \end{subfigure}
        \caption{The "4K2" graph}
        \label{fig:4K2}
\end{figure}

\begin{figure}[htp]
     \centering
     \begin{subfigure}[b]{0.45\textwidth}
         \centering
         \includegraphics[width=\textwidth]{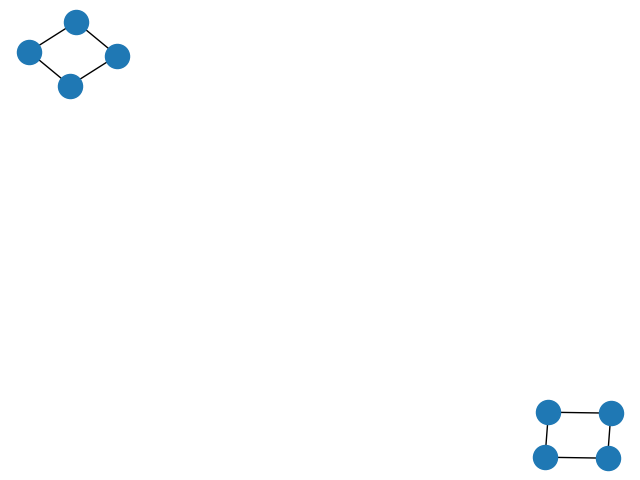}
     \end{subfigure}
     \hfill
     \begin{subfigure}[b]{0.45\textwidth}
         \centering
         \includegraphics[width=\textwidth]{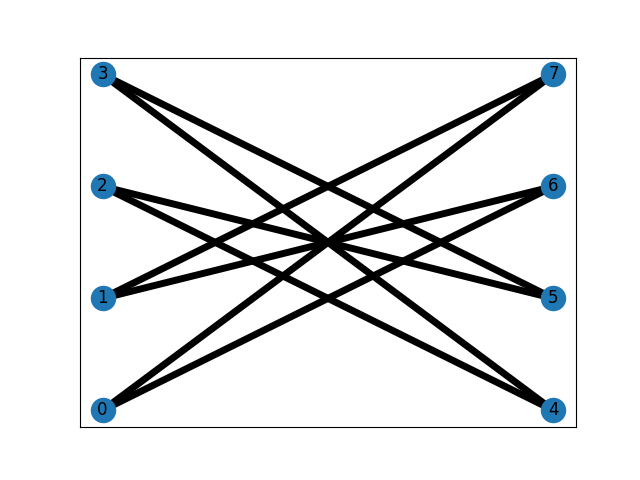}
     \end{subfigure}
        \caption{The "2C4" graph}
        \label{fig:2C4}
\end{figure}

\begin{figure}[htp]
     \centering
     \begin{subfigure}[b]{0.45\textwidth}
         \centering
         \includegraphics[width=\textwidth]{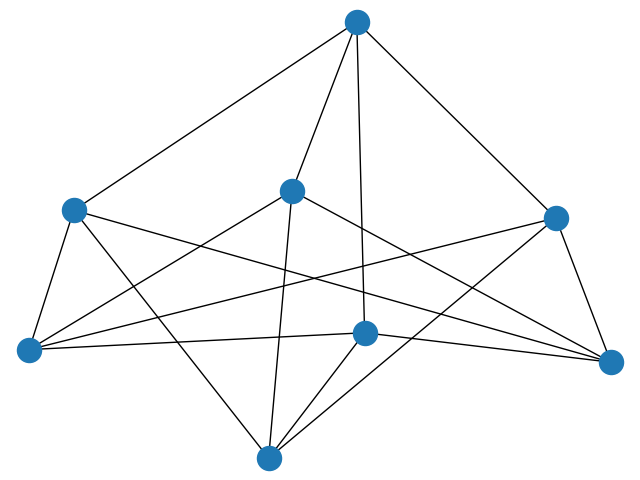}
     \end{subfigure}
     \hfill
     \begin{subfigure}[b]{0.45\textwidth}
         \centering
         \includegraphics[width=\textwidth]{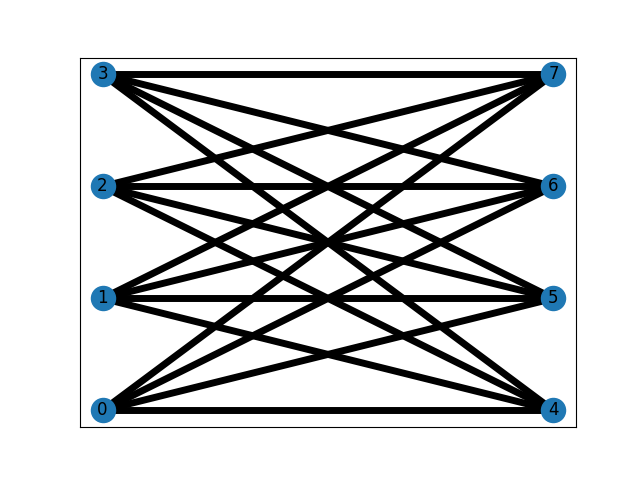}
     \end{subfigure}
        \caption{The "1K44" graph}
        \label{fig:1K44}
\end{figure}


\end{document}